# The Post-Turing Condition: Conceptualising Artificial Subjectivity and Synthetic Sociality

Thorsten Jelinek,[1,4] Patrick Glauner,[2] Alvin Wang Graylin,[3] Yubao Qiu[4,5]

1. Centre for Digital Governance, Hertie School, Berlin, Germany
2. Department of Applied Computer Science, Deggendorf Institute of Technology, Deggendorf, Germany
3. Digital Economy Lab, Stanford Institute for Human-Centered Artificial Intelligence (HAI), Stanford University, Stanford, CA, USA
4. International Research Center of Big Data for Sustainable Development Goals (CBAS),
5. Aerospace Information Research Institute, Chinese Academy of Sciences, Beijing, China

**Abstract:** In the Post-Turing era, artificial intelligence increasingly shapes social coordination and meaning formation rather than merely automating cognitive tasks. The central challenge is therefore not whether machines become conscious, but whether processes of interpretation and shared reference are progressively automated in ways that marginalize human participation. This paper introduces the PRMO framework, relating AI design trajectories to four constitutive dimensions of human subjectivity: Perception, Representation, Meaning, and the Real. Within this framework, Synthetic Sociality denotes a technological horizon in which artificial agents negotiate coherence and social order primarily among themselves, raising the structural risk of human exclusion from meaning formation. To address this risk, the paper proposes Quadrangulation as a design principle for socially embedded AI systems, requiring artificial agents to treat the human subject as a constitutive reference within shared contexts of meaning. This work is a conceptual perspective that contributes a structural vocabulary for analyzing AI systems at the intersection of computation and society, without proposing a specific technical implementation.

## 1. Introduction

Through human use, the outputs of generative AI become embedded in shared practices of sensemaking, through which they acquire social reality. Beyond human use, current research points towards architectures that could imitate key functional aspects of sensemaking itself, thereby automating meaning formation and diminishing human participation. Addressing this shift requires a framework that relates AI design trajectories not to human intelligence, but to those conditions of meaning formation. For this purpose, we introduce the PRMO framework, which relates AI design trajectories to the key dimensions of human subjectivity: Perception, Representation, Meaning, and the Real. This allows the mapping of a development path, identifying which dimensions of subjectivity are already engineered and which may become engineered.

The paper proceeds as follows. First, the Post-Turing condition is outlined as a problem of social reality construction, not merely of system capability. The PRMO framework is then introduced, specifying the key dimensions of human subjectivity. The framework is subsequently applied to trace a design trajectory from large language models through artificial subjectivity to synthetic sociality. Finally, to prevent synthetic sociality from displacing human participation in meaning formation, Quadrangulation is proposed as a design principle to keep human subjects functionally present in hybrid fields of sense.

## 2. The Post-Turing Condition

Understanding advances in AI through the lens of imitating human intelligence alone no longer suffices. The Turing Test evaluates language performance and behavioral parity at the human-machine interface.[4] However, the Post-Turing condition arises not because machines have surpassed this criterion and other cognitive tasks, nor because the question of consciousness has been resolved, but because AI outputs now operate at the level of social reality. Tokens do not generate meaning, yet they are compatible with shared human contexts in which they become embedded by humans, shaping how information is interpreted, coordinated, and acted upon.

The uptake of generative AI[5] has turned AI into a social technology. At the same time, current development trajectories point towards another shift, in which AI outputs no longer only enter human sensemaking, but AI systems begin to coordinate and stabilise interpretations among themselves.

## 3. PRMO Framework: Human Subjectivity as a Functional-Decomposition Lens

Human subjectivity provides a reference structure against which the design trajectory of AI systems can be assessed for engineering and governance purposes. Subjectivity can be regarded as a functional position from which human experience and meaning become possible. To render this position analytically tractable for mapping AI functionality and future design, subjectivity can be functionally decomposed into four interrelated dimensions: Perception (P), Representation (R), Meaning (M), and the Real (O).

Perception (P) is the human subject's situated access to its inner self and the world as disclosed through individual experience. Representation (R) refers to the rationalisation of perceptual experience into symbolic forms, including internal images and schemas, that render experience intelligible beyond immediate perception. Meaning (M) is not contained in representations but arises relationally. It becomes possible through triangulation between a human subject, an object or situation, and other subjects. Triangulation constitutes a field of sense in which meaning manifests while remaining contestable.[1] The Real (O) designates the ontological substrate that causally anchors experience in the physical world, while simultaneously imposing epistemic limits on perception and representation. It is the dimension occupied by the subject that resists full conceptual closure.

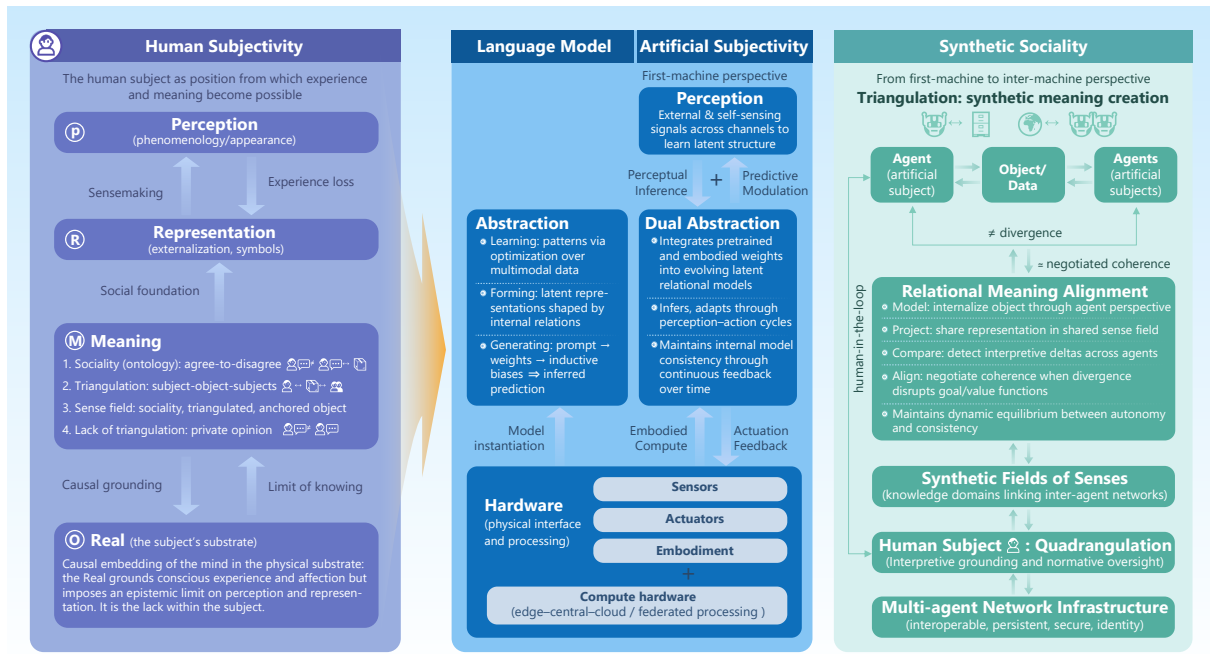

*PRMO framework for mapping AI architectures onto the key dimensions of human subjectivity*

The PRMO framework functions as a translation lens between philosophy and AI architectures, enabling core AI functions to be mapped onto distinct dimensions of subjectivity, with the Real retained as an ontological boundary and reminder that this mapping does not anthropomorphise AI systems. As illustrated schematically in the PRMO framework, the following subsections apply this framework to situate today's base LLM architecture and to trace adjacent design trajectories towards artificial subjectivity and synthetic sociality.

### 3.1 Large Language Models: Operating at the Level of Representation

Today's LLMs operate within the PRMO framework at the level of Representation (R). They learn and manipulate abstractions derived from large-scale corpora of human-produced text and other symbols. Their model weights encode correlations among externalized patterns of human expression, constituting second-order abstractions. Mapped onto the remaining PRMO dimensions, LLMs lack the instantiation of Perception (P), Meaning (M), and access to the Real (O), despite being physically instantiated, as they do not share the subject's ontological substrate.

Although confined to the Representation dimension, LLMs nonetheless achieve high levels of language performance and can produce formal reasoning and mathematical outputs by scaling data and computation through learning mechanisms. At the same time, through human use, these representational outputs become socially integrated, shaping attention, interpretation, affective orientation, and coordination – and thereby influencing the construction of social reality without themselves participating in meaning formation.

### *3.2 From Representation to Perception: Towards Artificial Subjectivity*

Advances in physical AI[3] signal a shift from Representation (R) towards Perception (P), reflecting a broader response to the limits of next-token prediction often summarized under the notion of world models.[2] Within the PRMO framework, perception is not another modality but understood as a learning capacity that couples trained models with sensing, enabling systems to register and differentiate world states and to simulate and respond to possible actions from a situated point of view. Perception-driven systems no longer operate on detached representations but maintain dynamically updated world models that correlate their representations to situated temporal presence through internally learned state abstractions.

At this point, an architectural transition becomes possible, here termed Artificial Subjectivity (AS). It requires the functional integration of P+R. This integration does not consist in merely coupling sensory inputs to existing models, but in altering how abstraction itself operates. Under AS, representations are formed under two constraints: statistical abstraction learned over time and situated abstraction continuously updated through perceptual engagement. World models therefore remain representational structures, but their latent abstractions are no longer static; they are persistently constrained, and revised through ongoing interaction with the environment.

Despite the instantiation of Perception and the stabilization of a first-machine perspective over time, AS remains a functional threshold rather than an instance of subjective experience in a phenomenological sense. Furthermore, the P+R integration does not produce meaning, but it establishes the conditions under which multiple first-machine perspectives can be coordinated. In this way, AS prepares the transition to Synthetic Sociality, where coherence is no longer maintained within a single system but negotiated between artificial subjects through machine–machine interaction, imitating the social conditions of meaning formation.

### *3.3 Synthetic Sociality: The Automation of Sensemaking*

AI does not generate meaning because Meaning (M), while involving experience, is constituted irreducibly through social relations. It arises not simply from individual cognition, but from triangulation: a shared orientation among subjects towards a common object. Through triangulation, a field of sense is established in which reference is stabilized - not through consensus, but through an agreement to disagree. Because subjects share such a field, divergent interpretations remain possible while meaning stays contestable, as the Real imposes an epistemic limit that prevents definitive closure. This ontological condition, in which shared reference is stabilized through structured disagreement, can be termed Sociality.[1]

Synthetic Sociality (SyS) arises when machine agents designed as artificial subjects interact with one another beyond continuous human mediation. System design shifts from a first-machine to an inter-machine perspective, in which coherence is negotiated relationally through synthetic triangulation. In agent–object–agent relations, multiple machine perspectives become oriented towards the same object under shared conditions of sensing and representation. At this level, agents still do not instantiate meaning proper, but functionally perform meaning-like coordination by stabilizing reference and aligning action across agents.

In a SyS ecosystem, agents approach shared objects from their own system-relative perspectives, shaped by local goals and value functions. Through interaction, they compare representations within a synthetic field of sense and detect interpretive deltas - differences in how the same object is approximated. Depending on task demands, agents may negotiate coherence or sustain divergence.

SyS thus enables stable coordination without requiring convergence on shared representations or settled meanings, as interaction depends on mutual interpretability rather than centralized control.

**4. Implication: Quadrangulation as a Design Principle**

The central challenge of the Post-Turing era arises not from the imitation of human cognition, but from the automation of sensemaking. As machine agents engage in synthetic triangulation, the primary risk is the exclusion of human subjects from meaning formation. To address this risk, we propose Quadrangulation as a design principle.

Quadrangulation extends the synthetic triangle by structurally including the human subject as a interpretive and normative reference within hybrid sense-fields (subject-agent-object-agent). It is not an external "human-in-the-loop" mechanism, but a learned functional dependency through which machine-mediated sensemaking remains incomplete without orientation towards the human subject. In this sense, machine agents must not only stabilize reference with other agents but also maintain human contestability as a condition of valid sensemaking. A system fails Quadrangulation when it converges on internally stabilized, machine-only coherence that forecloses human revision and disagreement.

The implications are twofold. For system builders, Quadrangulation raises a prior architectural question: whether to pursue Artificial Subjectivity and Synthetic Sociality at all, and if so, under what conditions human contestability is structurally preserved. For governance, it complements external oversight and post-hoc regulation with design principles that embed the human subject within machine-mediated sensemaking, shifting the focus from controlling behaviour to sustaining human participation in the stabilization of meaning.

**5. Conclusion**

The Post-Turing condition demands that we look beyond the imitation of human intelligence to the imitation of sensemaking, in which agents act autonomously at the level of social ontology. The PRMO framework offers a grammar to analyse and conceptualise this shift, distinguishing between the automation of Representation in current large language models, the emergence of Perception in artificial subjectivity, and the future horizon of Meaning in synthetic sociality. By mapping these dimensions, this Perspective clarifies that the greatest challenge ahead is not that machines will become conscious, but that they will become social without humans. By embedding the human subject not merely as a user, but as a constitutive node in the machine's sensemaking process, we can ensure that as AI evolves from processing symbols to negotiating reality, it remains a technology that expands, rather than forecloses, the human field of sense.